\documentclass[twocolumn,showpacs,amsmath,pre]{revtex4}

\usepackage{graphicx}
\usepackage{amsmath}
\usepackage{amssymb}
\usepackage{natbib}

\begin{document}

\title{Jammed particulate systems are inherently nonharmonic}

\author{Carl F. Schreck$^{1}$}
\author{Thibault Bertrand$^{2}$}
\author{Corey S. O'Hern$^{3,1}$}
\author{M. D. Shattuck$^{4}$} 

\affiliation{$^{1}$Department of Physics, Yale University, New Haven,
  Connecticut 06520-8120, USA}

\affiliation{$^{2}$D\'{e}partement Physique, Ecole Normale Sup\'{e}rieure de 
Cachan, 61 Avenue du Pr\'{e}sident Wilson, 94235 Cachan, France}

\affiliation{$^{3}$Department of Mechanical Engineering and Materials Science, Yale University, New
  Haven, Connecticut 06520-8286, USA}

\affiliation{$^{4}$Benjamin Levich Institute and Physics Department, The 
City College of the City University of New York, New York, New York 10031, USA}

\begin{abstract}
Jammed particulate systems, such as granular media, colloids, and
foams, interact via one-sided forces that are nonzero only when
particles overlap.  We find that systems with one-sided repulsive
interactions possess no linear response regime in the large system
limit ($N\rightarrow \infty$) for all pressures $p$ (or compressions
$\Delta \phi$), and for all $N$ near jamming onset $p\rightarrow 0$.
We perform simulations on 2D frictionless bidisperse mechanically
stable disk packings over a range of packing fractions $\Delta \phi =
\phi-\phi_J$ above jamming onset $\phi_J$.  We apply perturbations
with amplitude $\delta$ to the packings along each eigen-direction
from the dynamical matrix and determine whether the response of the
system evolving at constant energy remains in the original eigenmode
of the perturbation.  For $\delta > \delta_c$, which we calculate
analytically, a single contact breaks and fluctuations abruptly spread
to all harmonic modes. As $\delta$ increases further all discrete
harmonic modes disappear into a continuous frequency band.  We find
that $\langle \delta_c \rangle \sim \Delta \phi/N^{\lambda}$, where
$1 > \lambda > 0.5$, and thus jammed particulate systems are
inherently nonharmonic with no linear vibrational response regime as
$N\rightarrow \infty$ over the full range of $\Delta \phi$, and as
$\Delta \phi \rightarrow 0$ at any $N$.
\end{abstract}


\maketitle

{\bf Introduction} Granular materials, which are collections of 
macroscopic grains that interact via contact forces, such as sand,
powders, pharmaceutical, and consumer products, display strongly
nonlinear spatio-temporal dynamics even when they are weakly driven.
In stark contrast to conventional solids~\cite{tencate} granular
solids possess nonaffine, hysteretic, and time-dependent mechanical
response~\cite{jacob}, and dispersive, attenuated, and noisy acoustic
response~\cite{liu,mouraille} for micro-strains.

Crystalline and amorphous atomic and molecular solids display
well-defined linear response regimes for small perturbations.
Similarly, there has been a large research effort to identify linear
response regimes for granular and other jammed particulate systems.
Examples include effective medium theory~\cite{makse} for granular
media, which provides predictions for the elastic moduli as a function
applied pressure, and approaches that assume the vibrational modes of
static, mechanically stable (MS) packings obtained from the dynamical
matrix in the harmonic approximation describe the mechanical
response~\cite{ohernJ}, vibrations~\cite{silbert1,yodh}, and heat
flow~\cite{vitelli} of weakly perturbed and fluctuating particulate
systems.

However, it has not been determined whether jammed
particulate systems possess a linear response regime, and if so, over
what range of perturbation amplitudes and timescales.  To address this
fundamental question for granular media, it is important to understand
separately the manifold contributions to nonharmonicity including
nonlinear, dissipative, and frictional particle
interactions~\cite{johnson}, inhomogeneous force
propagation~\cite{force_chain,owens}, and breaking and forming of
intergrain contacts~\cite{liu}.  Here, we describe
computational studies to quantify perhaps the most important
contribution to nonharmonicity in jammed particulate media---the
one-sided nature of contact interactions---interparticle forces are
only nonzero when two grains are in contact, but are strictly zero when
they are out of contact.

We find that one-sided interactions make jammed particulate materials
{\it inherently} nonharmonic, {\it i.e.} nonharmonic even in the limit
of vanishing perturbation amplitude, due to changes in the contact
network following the perturbation~\cite{tournat}.  Specifically, we
employ the harmonic approximation and calculate the eigenmodes of the
dynamical matrix~\cite{barrat} for MS frictionless packings, subject
the packings to vibrations along the harmonic set of eigenmodes, and
quantify the frequency content of the response versus the perturbation
amplitude $\delta$.  We find that systems become nonharmonic ({\it
i.e.} the response is not confined to the original mode of excitation)
when only a {\it single} contact is broken (or gained) at a critical
$\delta_c$ that depends on the original mode of excitation.  For
$\delta > \delta_c$ the response first spreads to all (harmonic)
eigenmodes with an amplitude that scales inversely with frequency, and
then becomes continuous with an average frequency that decreases with
$\delta$.  We show that $\langle \delta_c \rangle$ averaged over the modes
of excitation tends to zero in the large system limit even for highly
compressed systems, and tends to zero in the limit of zero compression
at all system sizes.  Thus, jammed particulate systems possess no
harmonic regime in the large system limit and at jamming onset for any
system size.

\begin{figure}
\includegraphics[width=.8\columnwidth]{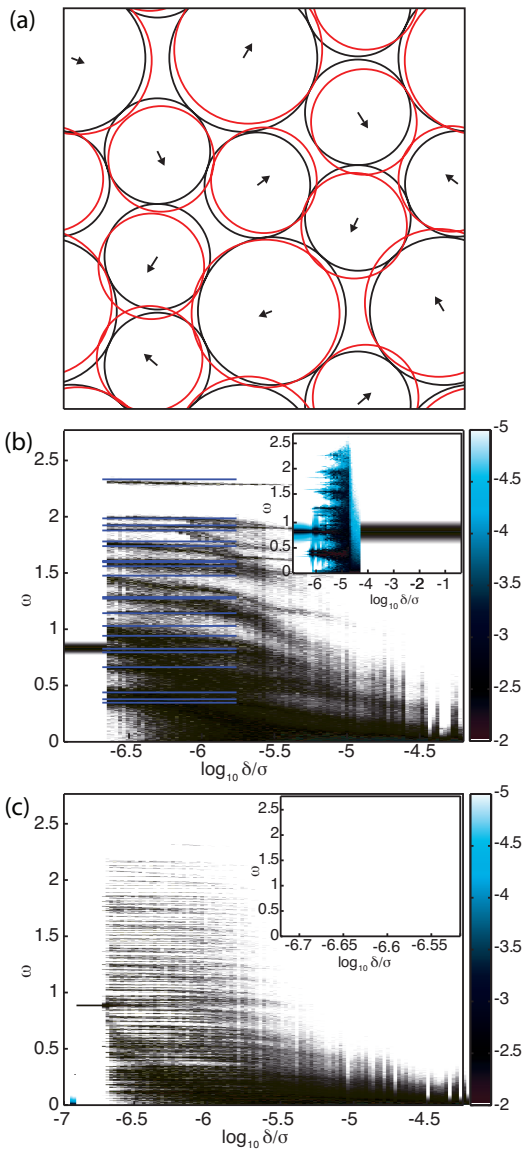}
\vspace{-0.1in}
\caption{(a) Mechanically stable (MS) packing of frictionless disks
for $N=12$ at $\Delta \phi=10^{-5}$ (black solid) and a packing
perturbed along the $6$th eigenmode of the dynamical matrix by $\delta
= 0.1 \sigma$ (red dashed). The vector lengths are proportional to the
displacements.  (b) An intensity plot of the logarithm of the
power spectrum $|{\vec R}(\omega)|^2$ as a function of
frequency $\omega$ and perturbation $\delta$ along the $6$th eigenmode
of the system in (a) after $170$ oscillations.  The solid
horizontal lines indicate the $22$ harmonic eigenfrequencies for
(a). The inset shows the same calculation except for a two-sided
linear spring potential. (c) Same as (b) except for $N=58$ at $\Delta
\phi=10^{-5}$ with perturbation in mode $40$ after $150$ oscillations.
The inset shows a close-up of the transition.}
\vspace{-0.1in}
\label{fig1}
\end{figure}

{\bf Model and Simulations} We focus on frictionless MS packings of
bidisperse disks in 2D with system sizes in the range $N=12$ to $1920$
particles using periodic boundaries in square simulation cells ($2N/3$
disks with diameter $\sigma$ and $N/3$ disks diameter $1.4\sigma$).
The disks interact via the linear repulsive spring potential
\begin{equation}
\label{potential}
V(r_{ij}) = \frac{\epsilon}{2} \left(1 -
\frac{r_{ij}}{\sigma_{ij}} \right)^2 \Theta \left(1 - \frac{r_{ij}}
{\sigma_{ij}} \right), 
\end{equation}
where $r_{ij}$ is the center-to-center separation between disks $i$
and $j$, $\epsilon$ is the characteristic energy scale, $\Theta(x)$ is
the Heaviside function, and $\sigma_{ij} = (\sigma_i+\sigma_j)/2$ is
the average diameter.  We have also studied systems with Hertzian and
purely repulsive Lennard-Jones interactions, but the repulsive linear
spring potential provides a `lower bound' on the degree of
nonlinearity arising from one-sided interactions. Energy, length, and
timescales are measured in units of $\epsilon$, $\sigma$, and
$\sqrt{m/\epsilon}\sigma$, respectively.

\begin{figure*}
\includegraphics[width=\textwidth]{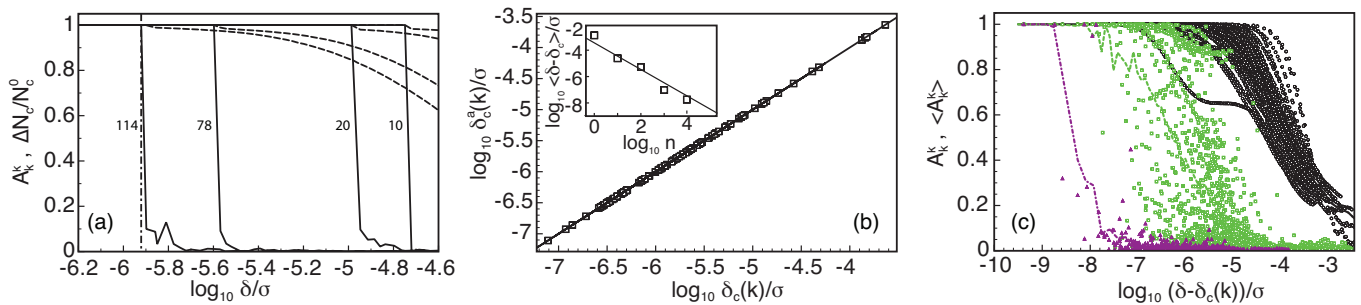}
\vspace{-1.4in}
\caption{(a) Amplitude $A_k^k$ (solid) and deviation of the number of
contacts $\Delta N_c=N_c^0 - \langle N_c \rangle_t$ relative to the
unperturbed number $N_c^{0}$ (dashed) versus perturbation amplitude
$\delta$ along four eigenmodes (labeled by mode number) for the system
in Fig.~\ref{fig1} (c) after $n=10^4$ oscillations.  The vertical
dot-dahsed line indicates $\delta = \delta^a_c$ for $k=114$. (b) The
measured $\delta^a_c(k)$ at which $A_k^k$ and $\langle N_c \rangle_t$
begin to deviate from $1$ for perturbations along all eigenmodes at $n
= 10^{4}$, $N=58$, and $\Delta \phi=10^{-5}$ versus the calculated
deformation amplitude $\delta_c(k)$ in Eq.~\ref{break} at which the
first contact breaks. The inset displays the values $\delta^* -
\delta_c(k)$ at which $\langle A_k^k\rangle$ decays to $0.2$ as a
function of $n$.  The solid line has slope $-1$. (c) $A_k^k$ versus
$\delta - \delta_c(k)$ for each $k$ (open symbols) and $\langle A_k^k
\rangle$ averaged over $k$ (lines) for $n = 1$ (circles, solid line),
$10^2$ (squares, dashed line), and $10^4$ (triangles, dot-dashed line)
oscillations after the perturbation.}
\label{fig2}
\vspace{-0.1in}
\end{figure*}

The MS packings were generated using the compression and energy
minimization protocol described in Ref.~\cite{gao}.  Each MS packing
is characterized by a packing fraction $\phi_J$ above which the
potential energy $V$ and pressure $p$ of the system begins to increase
from zero. The distance in packing fraction from $\phi_J$ is tuned
from $\Delta \phi = 10^{-8}$ to $10^{-1}$ and the positions of the
particles are accurate to $10^{-16}$ at each $\Delta \phi$.  We
calculate the eigenfrequencies $\omega_i$ and eigenmodes ${\hat e}_i =
\{ {\vec e}_i^1, {\vec e}_i^2,\ldots,{\vec e}_i^N \} = \{ e_{xi}^1,
e_{yi}^1,e_{xi}^2, e_{yi}^2,\ldots,e_{xi}^N,e_{yi}^N \}$ (with ${\hat
e}_i^2 = 1$) in the harmonic approximation from the dynamical matrix
evaluated at the MS packing.  Since the systems are mechanically
stable, the ${\cal N} = 2N' - 2$ eigenfrequencies $\omega_i >
0$~\cite{supplement}, where $N' = N - N_r$ and $N_r$ is the number of
rattler particles with less than three contacts per particle.  We
index the eigenfrequencies from smallest to largest, $i=1$ to ${\cal
N}$, removing the two trivial eigenfrequencies corresponding to
uniform translations.

To test whether the packings possess a harmonic regime, we apply
displacements to individual particles and then evolve the system at
constant total energy $E$.  Specifically, at time $t=0$, we apply the
displacement
\begin{equation}
{\vec R} - {\vec R}^0 = \delta {\hat e}_i,
\end{equation}
where the new configuration ${\vec R} = \{ {\vec R}_1, {\vec
R}_2\ldots, {\vec R}_N \} = \{x_1, y_1, \ldots, x_N,y_N\}$, and ${\vec
R}^0$ is the original MS packing.  We remove rattlers from the MS
packings prior to applying the perturbations. A sample perturbation
for $N=12$ along the $6$th mode is shown in Fig.~\ref{fig1} (a). For
$t>0$, we solve Newton's equations of motion at constant $E$, and
measure the particle displacements and number of contacts as a
function of the number of oscillations $n$ for perturbations along
each mode $k$.

{\bf Results} In Fig.~\ref{fig1} (b), we show the
logarithm of the power spectrum $|{\vec R}(\omega)|^2$, where ${\vec
R}(\omega) = \int_0^{n T_6} dt e^{i \omega t} {\vec R}(t)$ for
$n=170$ oscillations, where $T_6 = 2 \pi/\omega_6$, as an intensity plot versus
the perturbation amplitude $\delta$ (along the $6$th mode) and
$\omega$ for the system shown in Fig.~\ref{fig1} (a) with linear
repulsive spring interactions.  This plot demonstrates several key
features: (1) There is an extremely sharp onset of nonharmonicity at
$\log_{10}~ \delta^a_c/\sigma \simeq -6.8$. For $\delta < \delta_c^a$, the
system vibrates with $\omega = \omega_6$. Although $\delta_c^a$ depends
on the excitation mode, the transition for each mode is sharp.  (2)
For $\delta \gtrsim \delta_c^a$, the response spreads to include other
harmonic eigenfrequencies (shown as solid horizontal lines in
Fig.~\ref{fig1} (b)) and $|{\vec R}(\omega)|^2 \sim \omega^{-2}$
similar to equipartition in thermal equilibrium. (3) For larger
perturbations, the power spectrum develops a continuous frequency band
in which the harmonic eigenfrequencies are completely lost.  At
sufficiently large amplitudes, the dominant contribution to the broad
power spectrum approaches $\omega = 0$.  Note that this crossover to 
nonharmonic frequency response occurs at much larger amplitudes in 
systems with smooth nonlinear interaction potentials.

For larger systems the transition from harmonic to nonharmonic
behavior is similar (Fig.~\ref{fig1} (c)).  The inset to
Fig.~\ref{fig1} (c) shows that large systems display an
intermediate nonharmonic regime in which a subset of harmonic
eigenmodes are populated at the onset of nonharmonicity, $\delta =
\delta_c^a$.  To put the effects of one-sided potentials into
perspective, we compare these results with those from two-sided spring
potentials ({\it i.e.} Eq.~\ref{potential} with the argument of
$\Theta$ replaced by $1-R_{ij}/\sigma_{ij}$).  For $N=12$, the
transition for systems with one-sided repulsive spring interactions
occurs at perturbations more than four orders of magnitude smaller
than those for systems with double-sided spring
potentials~\cite{supplement} and the transition occurs slowly over a
decade in $\delta$ (inset to Fig.~\ref{fig1} (b)).

To quantify the harmonic to nonharmonic transition, we calculate the
number of particle contacts $\langle N_c \rangle_t$ averaged over time
and define a harmonicity parameter $A_k^k$ that measures the spectral
content of the particle displacements in the eigenmode direction $k$
at eigenfrequency $\omega_k$ following a perturbation along eigenmode
$k$:
\begin{equation}
\label{p1b}
A_k^k = \left| \frac{\int_0^{nT_k}\Delta {\vec R}(t)\cdot {\hat e}_k \cos
(\omega_k t) dt }{\delta \int_0^{nT_k} \cos^2(\omega_k t) dt} \right|,
\end{equation}
where $\Delta {\vec R}(t) = {\vec R}(t) - \langle {\vec R}(t)
\rangle_t$.  $A_k^k=1$ for harmonic systems and $A^k_k \approx 0$ for
nonharmonic systems that do not oscillate in mode $k$ at
$\omega_k$. We also calculate the harmonicity parameter $\langle A_k^k
\rangle$ averaged over all individually perturbed modes $k$.

In Fig.~\ref{fig2} (a), we plot $A_k^k$ and the deviation in the
time-averaged number of contacts $\Delta N_c = N_c^0 - \langle N_c
\rangle_t$ relative to the unperturbed value $N_c^0$ versus $\delta$
along several modes $k$ for the system in Fig.~\ref{fig1} (c). We find
that $A_k^k$ for each mode $k$ begins to decrease from $1$ at the {\it
same} $\delta_c^a(k)$ where the average number of contacts $\langle
N_c\rangle_t$ begins to deviate from $N_c^0$.  For perturbations along
each mode $k$, the transition from harmonic to nonharmonic behavior
occurs when a {\em single} existing contact breaks.  To verify this,
we plot in Fig.~\ref{fig2} (b) $\delta^a_c(k)$ versus the predicted
amplitude $\delta_c(k)$ at which the first contact breaks.  The
predicted value $\delta_c(k)$ is obtained by solving $R^2_{ij} =
\sigma_{ij}^2$ for all contacting pairs of particles $i$ and $j$ for a
given MS packing and perturbation along mode $k$, and identifying the
minimum $\delta_c(k) = \min_{ij} |\delta_{ij}(k)|$, where
\begin{equation}   
\label{break}
\delta_{ij}(k) = \frac{ |{\vec e}_k^{ij}\cdot{\vec R}^0_{ij}| } {
  |{\vec e}_k^{ij}|^2 } \left( \sqrt{1+ \frac{(\sigma_{ij}^2 - |{\vec
      R}^0_{ij}|^2) |{\vec e}_k^{ij}|^2 }{|{\vec e}_k^{ij}\cdot{\vec
      R}^0_{ij}|^2} } -1 \right). 
\end{equation}
We find that the $\delta$ at which $A_k^k$ begins to decrease,
$\delta_c^a(k) = \delta_c(k)$ at which a single contact breaks (as
shown in Fig.~\ref{fig2} (b)) over a wide range of $\Delta \phi$ and
$N$ with a relative error less than $10^{-3}$ over four orders of
magnitude in $\delta_c(k)$.  For larger system sizes, it is possible
that new contacts can form before existing contacts break, but we find
that this does not occur for the system sizes and compressions
studied.

\begin{figure*}
\includegraphics[width=0.75\textwidth]{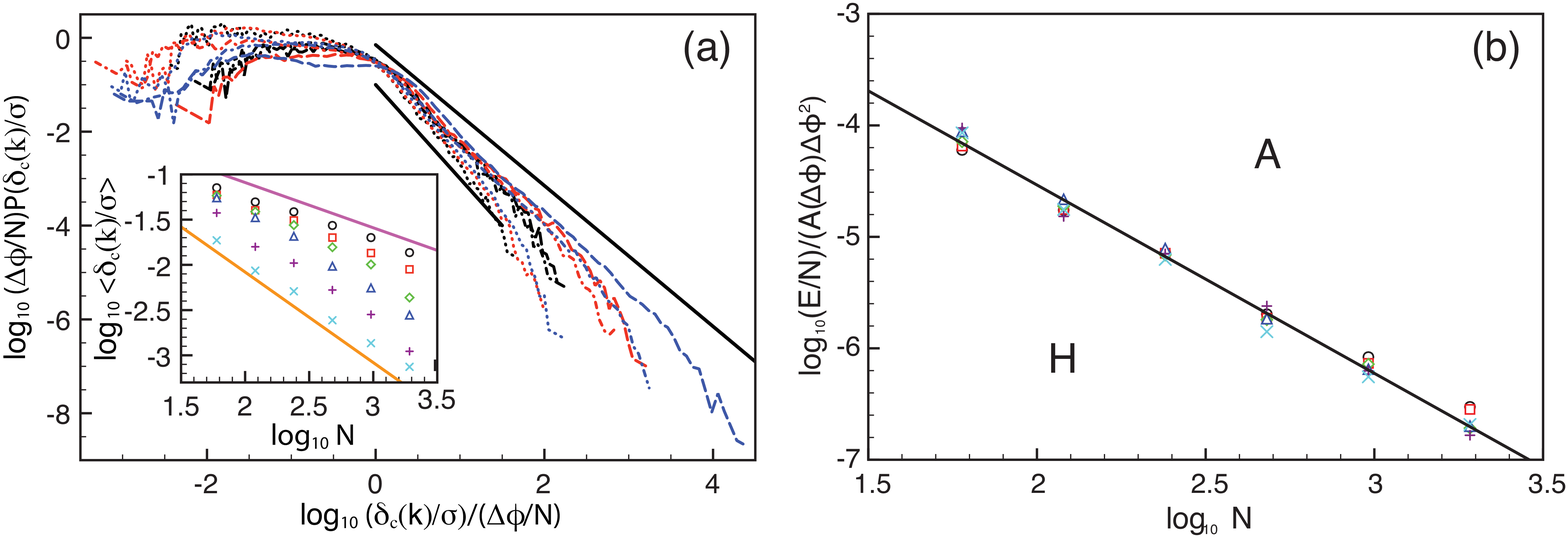}
\vspace{-0.15in}
\caption{(a) Distribution $P(\delta_c(k))$ (scaled by $\Delta \phi/N$)
versus $\delta_c(k) N/\Delta \phi$ for $\Delta \phi = 10^{-2}$
(dotted), $10^{-4}$ (dot-dashed), and $10^{-7}$ (dashed) and $N=60$
(black), $240$ (red), and $1920$ (blue).  The two solid black lines
have slope $1.5$ and $2$. The inset shows the scaling of $\langle
\delta_c(k) \rangle$ with $N$ for $\Delta \phi = 10^{-2}$ (crosses),
$10^{-3}$ (pluses), $10^{-4}$ (upward triangles), $10^{-5}$
(diamonds), $10^{-6}$ (squares), and $10^{-7}$ (circles).  The two
solid lines have slope $1$ and $0.5$.  (b) The total energy per
particle required to break a single contact averaged over $k$
(scaled by $A(\Delta \phi) (\Delta \phi)^2$) versus system size $N$
for $\Delta \phi = 10^{-2}$ (crosses), $10^{-3}$ (pluses), $10^{-4}$
(upward triangles), $10^{-5}$ (diamonds), $10^{-6}$ (squares), and
$10^{-7}$ (circles).  The solid line has slope $-1.7$.}
\label{fig3}
\vspace{-0.2in}
\end{figure*}

The rate at which energy input via a perturbation along eigenmode $k$
is transferred out of that mode and into other displacement modes
determines the shape of the decay of $A_k^k$.  In Fig.~\ref{fig2} (c),
we show $A_k^k$ and $\langle A_k^k \rangle$ versus $\delta -
\delta_c(k)$ for $n=1$, $10^2$, and $10^4$ oscillations for
perturbations along each mode $k$ individually.  For small $n$, even
though $A_k^k$ begins to decrease from $1$ at $\delta_c(k)$, the
shape of the decay depends on $k$ and the sharp decrease from $1$ to
$0$ occurs at small but finite $\delta -\delta_c(k)$. In the inset to
Fig.~\ref{fig2} (b), we measure the amplitude $\delta^*-\delta_c(k)$
at which $\langle A_k^k \rangle$ decays to a small value ($0.2$), and
find $\delta^*-\delta_c(k) \sim 1/n$.  For all $n$ and perturbations
$\delta$ studied in Fig.~\ref{fig2}, there is no detectable
nonharmonicity from the smooth nonlinearities in the
potential~\cite{supplement} in Eq.~\ref{potential}.

Thus, $\delta_c(k)$ is the critical deformation amplitude above
which MS packings become nonharmonic, and $\delta_c(k)$ can be
calculated exactly using Eq.~\ref{break} for each MS packing and mode
$k$.  In Fig.~\ref{fig3} (a) we show the distribution
$P(\delta_c(k)/\sigma)$ of critical amplitudes with the vertical and
horizontal axes scaled by $\Delta \phi/N$ to achieve approximate
collapse.  We find that the distribution $P(\delta_c(k)/\sigma)$
scales roughly as a power law $(\delta_c(k)/\sigma)^{\alpha}$ for
large $\delta_c$ with $\alpha \approx 2$ ($1.5$) for large (small)
$\Delta \phi$.  The distribution is cut-off and remains nearly
constant for $\delta_c(k)/\sigma < \Delta \phi/N$.  In the inset to
Fig.~\ref{fig3} (a), we plot $\langle \delta_c(k) \rangle$ averaged
over $k$ versus $N$ over a range of $\Delta \phi = 10^{-7}$ to
$10^{-2}$.  As expected from the power-law distribution in the main
plot, $\langle \delta_c(k) \rangle \sim N^{\alpha-1}$.  For all
$\Delta \phi$, the critical deformation amplitude scales to zero in
the large system limit.

The potential energy $V$ of a MS packing prepared at $\Delta \phi$ is
given by $V/N = B (\Delta \phi)^2$~\cite{ohernJ}, where $B$ is a
$O(1)$ constant.  We find that the average deformation energy $E^*
\approx \langle (\omega_k \delta_c(k))^2 \rangle$ for the critical
amplitude scales as $E^* \sim A(\Delta \phi) (\Delta
\phi)^2/N^{\beta}$, where $A(\Delta \phi)$ is only weakly dependent on
$\Delta \phi$ and $\beta \approx 1.7$.  For $E>E^*$ (labeled $A$ in
Fig.~\ref{fig3} (b)), MS packings are strongly anharmonic.  MS
packings are only harmonic for $E < E^*$, where $E^* \rightarrow 0$ in
the large system limit for all $\Delta \phi$.  Thus, eigenfrequencies
of the dynamical matrix do not describe vibrations of MS packings as
$N\rightarrow \infty$ for all $\Delta \phi$.

{\bf Conclusion} We have shown that one-sided repulsive
interactions in jammed particulate systems make them inherently
nonharmonic.  In the large system limit at any compression and in the
$\Delta \phi \rightarrow 0$ limit at any system size, infinitesimal
perturbations will cause them to become strongly nonharmonic, which
will affect their mechanical response, specific heat, and
energy diffusivity. In future studies, we will explore the possibility
of defining dynamic steady-state packings with robust effective harmonic modes
obtained from average particle positions.

{\bf Acknowledgments}
This research was supported by the National Science Foundation under
Grant Nos. DMS-0835742 (CO, TB, CS) and CBET-0968013 (MS).

\end{document}